\documentclass{article}
  \usepackage{cite}
  \usepackage[pdftex]{graphicx}
\usepackage{amsmath}
  \usepackage[caption=false,font=footnotesize,labelfont=sf,textfont=sf]{subfig}
\usepackage{url}

\usepackage{enumitem}
\usepackage[super]{nth}
\usepackage{multirow}
\usepackage{algorithm}
\usepackage[noend]{algorithmic}
\usepackage{booktabs}
\usepackage{soul} 
\usepackage{comment}
\usepackage{authblk} 

\usepackage{xcolor}

\usepackage{etoolbox}
\apptocmd{\thebibliography}{\setlength{\itemsep}{0pt}}{}{}

\begin{document}
%
\title{Who's in the Gang? Revealing Coordinating Communities in Social Media}

\author[1,2]{Derek Weber}
\author[1]{Frank Neumann}
\affil[1]{School of Computer Science, University of Adelaide, Australia}
\affil[2]{Defence Science and Technology Group, Adelaide, Australia}


%


\maketitle

\begin{abstract}
Political astroturfing and organised trolling are 
online malicious behaviours with significant real-world effects. 
Common approaches examining these phenomena focus on 
broad campaigns rather than the small groups responsible. 
To reveal latent networks of cooperating accounts, we propose a novel temporal window 
approach that relies on account interactions and metadata alone. 
It detects groups of accounts engaging in  
behaviours that, in concert, 
execute different goal-based strategies, 
which we describe. 
Our approach is validated against two relevant datasets with ground truth data.
\end{abstract}


%

\section{Introduction}\label{sec:intro}

Online social networks (OSNs) 
have established themselves 
as flexible and accessible systems for activity coordination and information dissemination.
This benefit was illustrated during the Arab Spring~\cite{carvin2013} but its danger continues in ongoing political interference~\cite{BessiFerrara2016,ferrara2017frelec,KellerICWSM2017}. 
Modern information campaigns are participatory, using the audience to amplify the desired narrative
~\cite{StarbirdAW2019cscw}. Through cyclical reporting, social media users can unknowingly become ``unwitting agents'' as ``sincere activists'' of state-based operations~\cite{StarbirdWilson2020}. 
The use of \emph{political} bots to influence the framing and discussion of issues in the mainstream media (MSM) 
remains prevalent~\cite{DebateNightICWSM2018,woolley2017us,BessiFerrara2016}. 
This \emph{megaphone effect} requires coordinated action and 
a degree of regularity that may leave traces in the digital record.

Relevant research has focused on high level analyses of campaign detection and classification~\cite{LeeCCS2013campext,varol2017campaigndetection,Alizadeh2020}, the identification of botnets and other dissemination groups~\cite{vo2017,Gupta2019,woolley2017us}, and coordination at the community level~\cite{KumarHLJ2018conflict,HineOCKLSSB2017kekcucks}. 
Some have considered generalised approaches to social media analytics~\cite{rapid2018,weber2019coord,Pacheco2020arxiv}, 
but unanswered questions regarding the clarification of coordination strategies remain.


We present a new approach to detect groups engaging in potentially coordinated activities, revealed through anomalous levels of coincidental behaviour. Links in the groups are inferred from behaviours that, with intent, are used to execute a number of identifiable coordination strategies. We validate our new technique on various datasets and show it successfully identifies coordinating communities.



Our approach 
infers ties between accounts based 
on 
activity 
to construct \emph{latent connection networks} (LCNs), 
in which \emph{highly coordinating communities} (HCCs) are 
detected. We use a variant of \emph{focal structures analysis} (FSA)~\cite{Sen2016fsa} to do this. A window-based 
approach is used to enforce temporal constraints
.

Comparison of  
two relevant 
datasets
, including 
labeled ground truth, with a randomised dataset provides validation. 
These 
research questions guided our evaluation:

\begin{description}
    \item[{\small RQ1}] How can HCCs be found in an LCN?
    \item[{\small RQ2}] How do the discovered communities differ? 
    \item[{\small RQ3}] Are the 
    HCCs 
    internally or externally focused? 
    \item[{\small RQ4}] How consistent is the HCC messaging?
    
\end{description}

This paper provides an overview of relevant literature, followed by a discussion of online coordination strategies and their execution. Our approach is then explained, and its experimental validation is presented\footnote{See \url{https://github.com/weberdc/find_hccs} for code and data.}.

\subsection{Related Work}

Sociological studies of influence campaigns can reveal their intent and how they are conducted. Starbird \emph{et al.}~\cite{StarbirdAW2019cscw} highlight three kinds: \emph{orchestrated}, centrally controlled campaigns (e.g., paid teams~\cite{theagency2015,king_pan_roberts_2017}); \emph{cultivated} campaigns that infiltrate existing movements; and \emph{emergent} campaigns arising from shared ideology (e.g., groups around conspiracists). Though their strategies differ, they use the same online interactions as normal users, but their patterns differ.


Computer science has focused on detecting information operations on social media via automation~\cite{rise2016}, 
campaign detection~\cite{LeeCCS2013campext,CaoCLGC2015urlsh,varol2017campaigndetection,Alizadeh2020},  
temporal patterns~\cite{chavoshi2017}, 
and community detection~\cite{morstatter2018alt,vo2017,Gupta2019}. Other studies have explored how bots and humans interact in political settings~\cite{DebateNightICWSM2018,BessiFerrara2016}, 
including 
exploring 
how deeply embedded bots appear in the network and their degree of organisation
~\cite{woolley2017us}. 
There is, however, a research gap: the computer science study of the ``orchestrated activities'' of accounts in general, regardless of their degree of automation
~\cite{GrimmeAA2018perspectives,Alizadeh2020}
.

Though some studies have observed the existence of strategic behaviour in and between online groups (e.g.,~\cite{KellerICWSM2017,KumarHLJ2018conflict,HineOCKLSSB2017kekcucks}), the challenge of identifying a broad range of strategies and their underpinning execution methods remains. 

Inferring social networks from OSN data requires attendance to the temporal aspect to understand information (and influence) flow and degrees of activity~\cite{holme2012}. Real time processing of OSN posts can enable tracking narratives via text clusters~\cite{AssenmacherATG2020}, but to process networks requires graph streams~\cite{McGregor2014graphstreams} or window-based pipelines (e.g.,~\cite{weber2019coord,Pacheco2020arxiv}).

This work contributes to the identification of strategic coordination behaviours,  
along with a general technique to enable detection of groups using them.

\section{Coordination Strategies}


Online influence 
relies on two primary mechanisms: \emph{dissemination} 
and \emph{engagement}. 
For example, an investigation of social media activity following UK terrorist attacks in 2017\footnote{\url{https://crestresearch.ac.uk/resources/russian-influence-uk-terrorist-attacks/}} identified accounts promulgating contradictory narratives, inflaming racial tensions and simultaneously promoting tolerance to sow division. By engaging aggressively, the accounts drew in participants who then spread the message.

\textbf{Dissemination} aims to maximise audience, to convince through repeated exposure and, in the case of malicious use, to cause outrage, polarisation and confusion, or at least attract attention to distract from other content.


\textbf{Engagement} is a subset of dissemination that solicits a response. It relies on targeting individuals or communities through mentions, replies and the use of hashtags as well as rhetorical approaches that invite responses (e.g., inflammatory comments or, as present in the UK terrorist example above, pleas to highly popular accounts).



A number of online coordination strategies have been observed in the literature making use of both dissemination and engagement, including: 

\begin{enumerate}
    \item 
    \textit{Pollution}: flooding a community with repeated or objectionable content, causing the OSN to shut it down~\cite{nasim2018,HineOCKLSSB2017kekcucks};
    \item 
    \textit{Boost}: heavily reposting content to make it appear popular
    ~\cite{CaoCLGC2015urlsh,vo2017,Gupta2019};
    \item 
    \textit{Bully}: groups of individuals harassing another individual or community~\cite{BurgessMF2016gamergate,KumarHLJ2018conflict}; and
    \item \textit{Metadata Shuffling}: groups of accounts changing 
    metadata to 
    hide their identities~\cite{Mariconti2017,ferrara2017frelec}.
\end{enumerate}

Different behaviour primitives (e.g., Table~\ref{tab:primitives}) can be used to execute these strategies. Dissemination can be carried out by reposting, using hashtags, or mentioning highly connected individuals in the hope they spread a message further. Accounts doing this covertly will avoid direct connections, and thus inference is required for identification. 
Giglietto \emph{et al.}~\cite{Giglietto2020} propose detecting anomalous levels of coincidental URL use as a way to do this; we expand this approach to other interactions. 

Some strategies require more sophisticated detection: detecting bullying through \emph{dogpiling} 
(e.g., during the \#GamerGate incident~\cite{BurgessMF2016gamergate}) requires collection of (mostly) entire conversation trees, which, while trivial to obtain on forum-based sites (e.g., Facebook and Reddit), are difficult on stream-of-post sites (e.g., Twitter, Parler and Gab). 
Detecting metadata shuffling requires long term collection on broad issues to detect the same accounts being active in different contexts.
\section{Methodology}

The major OSNs share a number of features, primarily in how they permit users to interact. 
By focusing on these commonalities, it is possible to develop approaches that generalise across the OSNs that offer them. 

Traditional social network analysis relies on long-standing relationships between actors. On OSNs these are typically friend/follower relations. These are expensive to collect and quickly degrade in meaning if not followed with frequent activity. By focusing on active interactions, it is possible to understand not just who is interacting with whom, but to what degree. This provides a basis for constructing (or inferring) social networks, acknowledging they may be transitory.

LCNs are built from inferred links between accounts. Supporting criteria include retweeting the same tweet (\emph{co-retweet}), using the same hashtags (\emph{co-hashtag}) or URLs (\emph{co-URL}), mentioning the same accounts (\emph{co-mention}), or joining the same `conversation' (a tree of \emph{reply} chains with a common root tweet) (\emph{co-conv}).

\begin{table}
    \centering
    \caption{Social media interaction equivalents}
    \label{tab:primitives}
    \resizebox{\columnwidth}{!}{%
        \begin{tabular}{@{}l|llllll@{}}
            \toprule
            OSN      & POST  & REPOST    & REPLY       & MENTION    & TAG        & LIKE \\
            \midrule
            Twitter  & tweet & retweet   & reply tweet & @mention   & \#hashtags & favourite \\
            Facebook & post  & share     & comment     & mention    & \#hashtag  & reactions \\
            Tumblr   & post  & repost    & comment     & @mention   & \#tag      & heart \\
            Reddit   & post  & crosspost & comment     & /u/mention & subreddit  & up/down vote \\
            \bottomrule
        \end{tabular}
    } 
\end{table}

\subsection{The LCN / HCC Pipeline} 

\begin{figure}[t!]
    \centering
    \includegraphics[width=0.95\columnwidth]{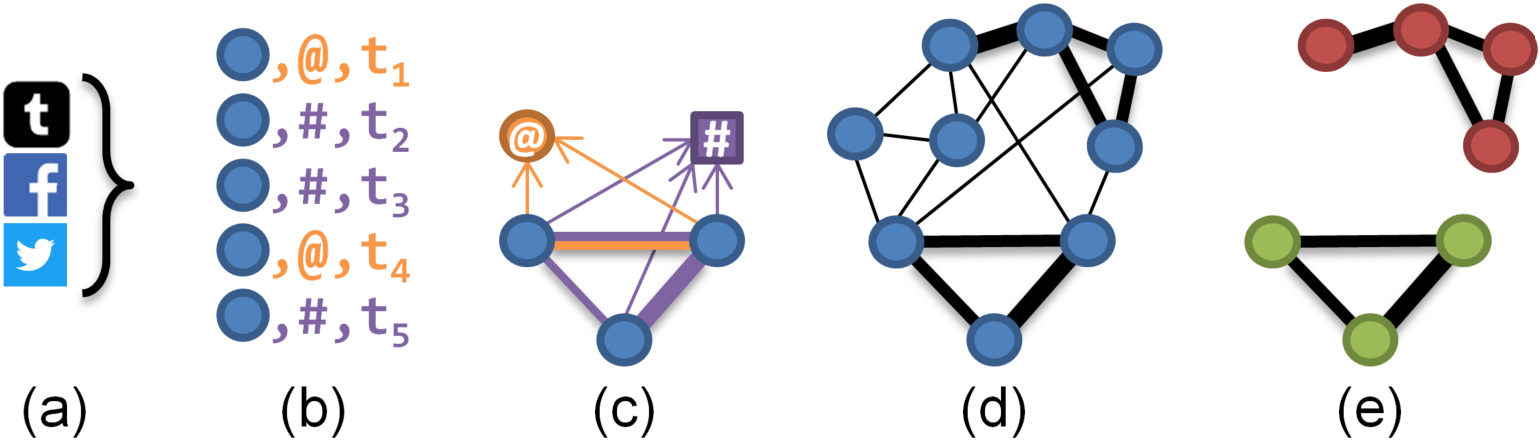}
    \caption{Conceptual LCN construction and HCC discovery process.}
    \label{fig:graph_construction}
\end{figure}

The key steps to 
extract HCCs from raw social media data are shown in Figure~\ref{fig:graph_construction}. 

\textbf{Step 1.} 
Convert social media posts $P$ 
to common interaction primitives, $I_{all}$. 
This step removes extraneous data and provides an opportunity for the fusion of sources. 


\textbf{Step 2.} 
From $I_{all}$, filter the interactions, $I_C$, relevant to the set $C$=$\{c_1, c_2, ..., c_q\}$ of criteria (e.g., co-mentions and co-hashtags). 
Illustrated in Figure~\ref{fig:graph_construction}b are the filtered mentions (in orange) and hashtag uses (in purple), ordered according to timestamp.


\textbf{Step 3.} 
Infer links between accounts 
given 
$C$, ensuring links are typed by criterion. 
The result, 
$M$, is a collection of inferred pairings. The count of inferred links between accounts $u$ and $v$ due to criterion $c \in C$ is $\beta^c_{\{u,v\}}$. Figure~\ref{fig:graph_construction}c shows inferred links between accounts 
due to common interactions.

\textbf{Step 4.} 
Construct an LCN, $L$, from the pairings in $M$. 
This network $L$=$(V,E)$ is a set of vertices $V$ representing accounts connected by weighted edges $E$ of inferred links. 
These edges represent evidence of different criteria linking the adjacent vertices. 
The weight of each edge $e_{\{u,v\}} \in E$ between vertices representing $u$ and $v$ 
for each criterion $c$ is $w^c_{\{u,v\}}$, and is equal to $\beta^c_{\{u,v\}}$.




Some community detection algorithms will require the multi-edges be collapsed to single edges, however, the edge weights are incomparable (e.g., retweeting the same tweet is not equivalent to using the same hashtag). For practical purposes, the inferred links can be collapsed 
and the weights combined for cluster detection using a simple summation, e.g., Equation~(\ref{eq:lcn_uv_combined_weight}), or a more complex process like varied criteria weighting. 
Implementations can retain information about how the edges were collapsed for later analysis, but the lack of multi-edges permits scope for more community detection algorithms to be used. 

\vspace{-0.5em}
\begin{equation} \label{eq:lcn_uv_combined_weight}
    w_{\{u,v\}} = \sum_{c=1}^{q} w^c_{\{u,v\}}
\end{equation}
\vspace{-0.5em}


Some criteria may result in highly connected LCNs, even if its members never directly interact. 
The final step filters out these coincidental connections. 


\textbf{Step 5.} 
Identify the highest coordinating communities, $H$, in $L$ (Figure~\ref{fig:graph_construction}e),  
using 
FSA\_V (Algorithm~\ref{alg:extract_hccs}), a variant of FSA~\cite{Sen2016fsa}, or an alternative community detection algorithm
, merging multi-edges as required. 
FSA\_V divides $L$ into communities using the Louvain algorithm~\cite{blondel2008} and builds candidate HCCs within each, starting with the `heaviest' (i.e., highest weight) edge (representing the most evidence of coordination). It then attaches the next heaviest edge until the candidate's mean edge weight (MEW) is no less than $\theta$ ($0 < \theta \leq 1$) of the previous candidate's MEW, or is less than $L$'s overall MEW. In testing, edge weights appeared to follow a power law, so $\theta$ was introduced to identify the point at which the edge weight drops significantly; $\theta$ requires tuning. A final filter ensures no HCC with a MEW less than $L$'s is returned. Unlike in FSA~\cite{Sen2016fsa}, recursion is not used, nor stitching of candidates, resulting in a simpler algorithm.

\begin{algorithm}[tb]
\caption{Extract HCCs (FSA\_V)}
\label{alg:extract_hccs}
\textbf{Input}: $L$=$(V,E)$: An LCN, $\theta$: HCC threshold \\
\textbf{Output}: $H$: Highly coordinating communities 
\begin{algorithmic}[1] 
    \STATE $E' \leftarrow \text{MergeMultiEdges}(E)$
    \STATE $g\_mean \leftarrow \text{MeanWeight}(E')$
    \STATE $louvain\_communities \leftarrow \text{ApplyLouvain}(L)$
    \STATE Create new list, $H$
    \FOR {$l \in louvain\_communities$}
        \STATE Create new community candidate, $h=(V_h,E_h)$
        \STATE Add heaviest edge $e \in l$ to $h$
        \STATE $growing \leftarrow$ \TRUE
        \WHILE{$growing$}
            \STATE Find heaviest edge $\vec{e} \in l$ connected to $h$ not in $h$
            \STATE $old\_mean \leftarrow \text{MeanWeight}(E_h)$
            \STATE $new\_mean \leftarrow \text{MeanWeight}(\text{Concatenate}(E_h, \vec{e}))$
            \IF{$new\_mean < g\_mean$ \OR \\ $new\_mean < (old\_mean \times \theta)$}
                \STATE $growing \leftarrow$ \FALSE
            \ELSE
                \STATE Add $\vec{e}$ to $h$
            \ENDIF
        \ENDWHILE
        \IF{MeanWeight$(E_h) > g\_mean$}
            \STATE Add $h$ to $H$
        \ENDIF
    \ENDFOR
\end{algorithmic}
\end{algorithm}

This algorithm prioritises edge weights while maintaining an awareness of the network topology by examining adjacent edges, something ignored by simple edge weight filtering. 
Our goal is to find sets of strongly coordinating users, so it is appropriate to prioritise strongly tied communities 
while still acknowledging coordination can also be achieved with weak ties (e.g., $100$ accounts paid to retweet a single tweet).

The complexity of the entire pipeline is low order polynomial, O($n^2$), due primarily to the pairwise comparison of accounts to infer links in Step $3$, which is constrained by window size when addressing the temporal aspect. 
Community detection algorithms designed for large networks may help to address this limitation~\cite{Fang2019}.


\subsection{Addressing the Temporal Aspect}


Temporal information is a key element of coordination
, and thus is critical for effective coordination detection. Frequent posts within a short period may represent genuine discussion or deliberate attempts to game trend algorithms~\cite{GrimmeAA2018perspectives,varol2017campaigndetection,AssenmacherATG2020}. 
We treat the post stream as a series of discrete windows to constrain detection periods
. An LCN is constructed from each window (Step 4), and these are then aggregated and mined for HCCs (Step 5). 
As we assume posts arrive in order, their timestamp metadata can be used to sort and assign them to windows.
\section{Evaluation and Validation} \label{sec:4exp}

Our approach was evaluated by searching for \emph{Boost} by co-retweet and other strategies in two datasets, while varying window sizes ($\gamma$). 
FSA\_V was compared against two other community detection algorithms when applied to the LCNs built in Step 4 (aggregated). 
We then validated the resulting HCCs through content, temporal and network analysis. 


\begin{table}[h]
    \centering
    \caption{Dataset statistics}
    \label{tab:dataset_stats}
    \resizebox{\columnwidth}{!}{%
    \begin{tabular}{@{}lrrrrrr@{}}
        \toprule
                       & Tweets (T) & \multicolumn{2}{c}{Retweets (RT)} & Accounts & T / Account / Day & RT / Account / Day  \\
                       \cmidrule{2-7}
        DS1            & 115,913   & 63,164   & (54.5\%)          & 20,563   & 0.31          & 0.17          \\
        - GT           & 4,193     & 2,505    & (59.7\%)          & 134      & 1.74          & 1.04          \\ 
        DS2            & 1,571,245 & 729,937  & (56.5\%)          & 1,381    & 3.12          & 1.45          \\ 
        \bottomrule
    \end{tabular}
    } 
\end{table}

\subsection{The Datasets}

The two real-world datasets selected (
Table~\ref{tab:dataset_stats}) represent two primary collection techniques: filtering a stream of posts using keywords direct from the OSN (DS1) and collecting the posts of specific accounts (DS2):

\begin{description}
    \item [{\small DS1}] Tweets relating to a regional Australian election in March 2018, including a ground truth subset (GT); and 
    \item [{\small DS2}] A large subset of the Internet Research Agency (IRA) dataset published by Twitter in October
    2018\footnote{\url{https://about.twitter.com/en_us/values/elections-integrity.html}}.
\end{description}

The data were collected, held and analysed in accordance with an approved ethics protocol\footnote{Protocol H-2018-045 was approved by the University of Adelaide's human research ethics committee.}.

DS1 was collected using 
RAPID~\cite{rapid2018} over an 18 day period (the election was on day 15) in March 2018. The filter terms included nine hashtags and $134$ political handles (candidate and party accounts)\footnote{Not included, but available on request, as per the ethics protocol.}. The dataset was expanded by retrieving all replied to, quoted and political account tweets posted during the collection period. The political account tweets formed our 
ground truth.


The IRA tweets cover 2009 to 2018, but DS2 consists of all posted in 2016, the year of the US Presidential election. 
Because DS2 consists entirely of IRA accounts~\cite{theagency2015}, 
it was expected to include evidence of cooperation.


\subsection{Set Up} \label{sec:setup}



Window size $\gamma$ was set at $\{15,60,360,1440\}$ (in minutes) and 
the three community detection methods used on the aggregated LCNs were:
\begin{itemize}
    \item FSA\_V ($\theta$=$0.3$);
    \item \textit{k nearest neighbour} (\textit{kNN}) with $k$=$ln(|V|)$ (\textit{cf.}~\cite{CaoCLGC2015urlsh}); 
    \item a simple threshold retaining the heaviest $90\%$ of edges.
\end{itemize}
Values 
for $\theta$ and the threshold were 
based on experimenting with values in $[0.1,0.9]$, maximising the MEW to HCC size ratio, using the $\gamma$=$\{15,1440\}$ DS1 and DS2 aggregated LCNs. 
Values for $\gamma$ were based on Zhao \emph{et al.}'s~\cite{DBLP:conf/kdd/ZhaoEHRL15} observation that 75\% of retweets occur within six hours of posting. This implies that if attempts were made to boost a tweet, retweeting it in much shorter times would be required for it to stand out from typical traffic.
Varol \emph{et al.}~\cite{varol2017campaigndetection} checked Twitter's trending hashtags every $10$ minutes, so values chosen for $\gamma$ ranged from 15m 
to a day, growing by a factor of approximately four at each increment. Coordinated retweeting was 
expected to occur in the smaller windows, but then replaced by coincidental co-retweeting as the window size increases. 


\subsection{Results}






The research questions introduced in Section~\ref{sec:intro} guide our discussion, but we also present follow-up analyses. 

\subsubsection{HCC Detection (RQ1)}

\textit{Detecting different strategies.}
The three detection methods all detected HCCs when searching for \emph{Boost} (co-retweets), \emph{Pollute} (co-hashtags), and \emph{Bully} (co-mentions) 
(Table~\ref{tab:strategies_info_1}). 
Notably, \textit{kNN} consistently builds a single large HCC, highlighting the need to filter the network prior to applying it (\textit{cf.}
~\cite{CaoCLGC2015urlsh}). The \textit{kNN} HCC is also consistently nearly as large as the original LCN for DS2, 
perhaps due to the low number of accounts and the fact that every edge of the retained vertices is retained, regardless of weight. It is not clear, then, that \textit{kNN} is producing meaningful results, even if it can extract a community.

\begin{table}[t]
    \caption{HCCs by coordination strategy}
    \label{tab:strategies_info_1}
    \resizebox{\columnwidth}{!}{%
    \begin{tabular}{@{}llr|rrr|rrr|rrr@{}}
        \toprule
        \multicolumn{3}{c}{}  & \multicolumn{3}{c}{GT} & \multicolumn{3}{c}{DS1} & \multicolumn{3}{c}{DS2} \\
        & Strategy & $\gamma$ & Nodes & Edges & Comp. & Nodes  & Edges     & Comp. & Nodes & Edges  & Comp. \\
        \midrule
        \multirow{3}{*}{\rotatebox[origin=c]{90}{LCN}} 
        & Boost    & 15       & 44    & 112   & 5    & 8,855  & 80,702    & 419  & 855   & 23,022 & 14 \\
        & Pollute  & 15       & 51    & 154   & 2    & 13,831 & 1,281,134 & 73   & 1,203 & 65,949 & 5 \\
        & Bully    & 60       & 70    & 482   & 1    & 16,519 & 1,925,487 & 222  & 1,103 & 37,368 & 5 \\
        \midrule
        \midrule
        \multirow{3}{*}{\rotatebox[origin=c]{90}{FSA\_V}} 
        & Boost    & 15       & 9     & 6     & 3    & 633   & 753     & 167  & 113   & 758   & 19 \\
        & Pollute  & 15       & 9     & 5     & 4    & 135   & 93      & 50   & 24    & 15    & 9 \\
        & Bully    & 60       & 11    & 7     & 4    & 338   & 280     & 119  & 109   & 1,123 & 16 \\
        \midrule
        \multirow{3}{*}{\rotatebox[origin=c]{90}{\textit{kNN}}} 
        & Boost    & 15       & 9     & 21    & 1    & 1,041 & 33,621  & 1    & 675   & 22,494 & 1 \\
        & Pollute  & 15       & 11    & 37    & 1    & 724   & 153,424 & 1    & 1,040 & 65,280 & 1 \\
        & Bully    & 60       & 18    & 135   & 1    & 1,713 & 663,413 & 1    & 692   & 35,136 & 1 \\
        \midrule
        \multirow{3}{*}{\rotatebox[origin=c]{90}{Threshold}} 
        & Boost    & 15       & 11    & 16    & 3    & 85    & 68     & 31   & 8     & 10    & 2 \\
        & Pollute  & 15       & 24    & 26    & 3    & 44    & 37     & 10   & 6     & 13    & 1 \\
        & Bully    & 60       & 15    & 19    & 3    & 25    & 23     & 8    & 10    & 10    & 3 \\
        \bottomrule
    \end{tabular}%
    }
\end{table}

\textit{Varying window size.} 
Different strategies may be executed over different time periods, based on their aims. \emph{Boost}ing a message to game trending algorithms requires the messages to appear close in time, whereas some forms of \emph{Bully}ing exhibit only consistency and low variation (mentioning the same account repeatedly). Polluting a user's timeline on Twitter can also be achieved by frequently joining their conversations over a sustained period. 
Varying $\gamma$ searching for \emph{Boost}, we found different accounts were prominent over different timeframes (Table~\ref{tab:hccs_info_1}); the overlap in the accounts detected in each timeframe differed considerably even though the number of HCCs stayed relatively similar. 
HCC sizes seemed to follow a power law; most were very small but a few were large.

\begin{table}[t]
    \caption{HCCs by window size $\gamma$ (Boost, FSA\_V)}
    \label{tab:hccs_info_1}
    \resizebox{\columnwidth}{!}{%
    \begin{tabular}{@{}lr|rrr|rr|rrrr@{}}
        \toprule
        \multicolumn{2}{c}{} & \multicolumn{3}{c}{Graph Attributes} & \multicolumn{2}{c}{HCC Sizes} & \multicolumn{4}{c}{Nodes in common} \\
        & $\gamma$        & Nodes  & Edges  & HCCs & Min. & Max. & $\gamma$=15 & $\gamma$=60 & $\gamma$=360 & $\gamma$=1440 \\
        \midrule
        \multirow{4}{*}{\rotatebox[origin=c]{90}{DS1}} 
        & 15   & 633 & 753 & 167 & 2 &  18 & 633 & 218 & 93 & 100           \\
        & 60   & 619 & 1,293 & 151 & 2 &  13 & - & 619 & 208 & 193           \\
        & 360  & 503 & 1,119 & 127 & 2 &  19 & - & - & 503 & 350           \\
        & 1440 & 815 & 2,019 & 141 & 2 &  110 & - & - & - & 815         \\
        \midrule
        \multirow{4}{*}{\rotatebox[origin=c]{90}{DS2}} 
        & 15   & 113 & 758 & 19 & 2 &  65 & 113 & 34 & 29 & 25           \\
        & 60   & 77 & 394 & 18 & 2 &  27 & - & 77 & 62 & 54           \\
        & 360  & 98 & 775 & 15 & 2 &  32 & - & - & 98 & 56            \\
        & 1440 & 69 & 380 & 15 & 2 &  27 & - & - & - & 69           \\
        \bottomrule
    \end{tabular}%
    }
\end{table}





\textit{HCC detection methods.}
Similarly, HCCs discovered by the three community extraction methods (Table~\ref{tab:hccs_info_2}) exhibit large discrepancies, suggesting that whichever method is used, tuning is required to produce interpretable results. 
This is evident in the literature: Cao \emph{et al.} conducted significant pre-processing when identifying URL sharing campaigns
~\cite{CaoCLGC2015urlsh}, and 
Pacheco \emph{et al.} showed how specific strategies could identify groups in the online narrative surrounding the Syrian White Helmet organisation~\cite{PachecoFM2020whitehelmets}.
Here we present the variation in results while controlling methods and other variables and 
keeping the coordination strategy constant, as our focus is the effectiveness of the method
.

\begin{table}[t]
    \caption{HCCs by detection method (Boost, $\gamma$=$15$) 
    }
    \label{tab:hccs_info_2}
    \resizebox{\columnwidth}{!}{%
    \begin{tabular}{@{}ll|rrr|rr|rrr@{}}
        \toprule
        \multicolumn{2}{c}{} & \multicolumn{3}{c}{Graph Attributes} & \multicolumn{2}{c}{HCC Sizes} & \multicolumn{3}{c}{Nodes in common} \\
        &               & Nodes  & Edges  & HCCs & Min. & Max. & FSA\_V & $kNN$ & Threshold \\
        \midrule
        \multirow{3}{*}{\rotatebox[origin=c]{90}{DS1}} 
        & FSA\_V    & 633 & 753 & 167 & 2 & 18 & 633 & 56 & 36        \\
        & kNN       & 1,041 & 33,621 & 1 & 1,041 & 1,041 & - & 1,041 & 44         \\
        & Threshold & 85 & 68 & 31 & 2 & 14 & - & - & 85        \\
        \midrule
        \multirow{3}{*}{\rotatebox[origin=c]{90}{DS2}} 
        & FSA\_V    & 113 & 758 & 19 & 2 & 65 & 113 & 88 & 4         \\
        & kNN       & 675 & 22,494 & 1 & 675 & 675 & - & 675 & 8         \\
        & Threshold & 8 & 10 & 2 & 2 & 6 & - & - & 8         \\
        \bottomrule
    \end{tabular}%
    }
\end{table}






\subsubsection{HCC Differentiation (RQ2)}



\textit{How similar are the discovered HCCs to each other and to the rest of the corpus?}
The HCC detection methods used relied on network information; in contrast we examine content, metadata and temporal information to validate the results. We contrast DS1 and DS2 results with GT (\emph{cf.}~\cite{KellerICWSM2017}) and a RANDOM dataset (\emph{cf.}~\cite{CaoCLGC2015urlsh}), constructed by randomly assigning non-HCC accounts from DS1 to groups matching the distribution of its HCCs (FSA\_V, $\gamma$=$15$). As DS2 consisted entirely of bad actors, it was felt non-HCC accounts from DS1 would be more representative of non-coordinating `normal' accounts.

\begin{figure}[t!]
    \centering
    \subfloat[GT.]{
        \includegraphics[width=0.41\columnwidth]{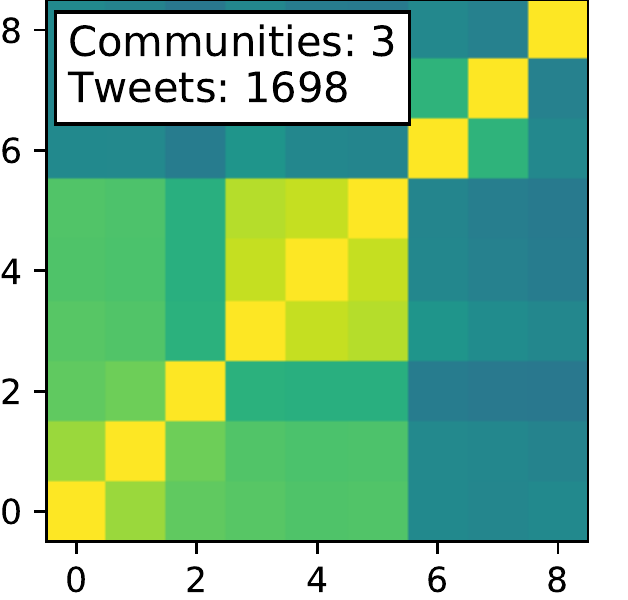}%
        \label{fig:sapol_sim_mtx_15m}
    } 
    \subfloat[DS1.]{
        \includegraphics[width=0.435\columnwidth]{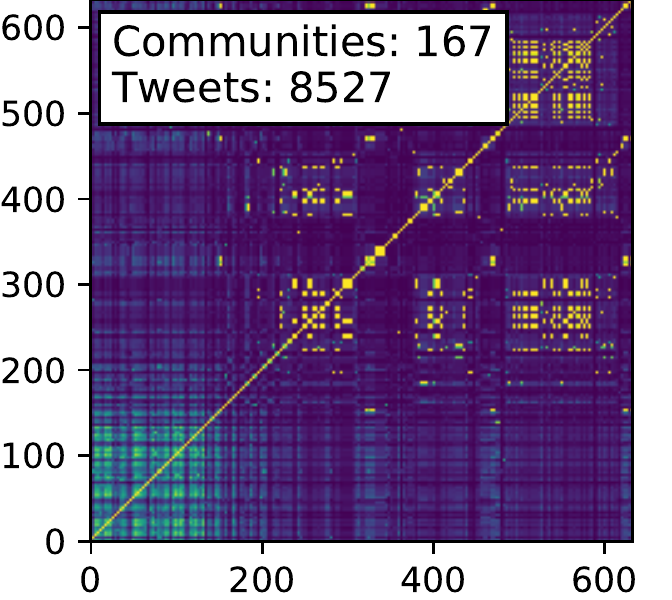}%
        \label{fig:saelec_sim_mtx_15m}
    } \\ 
    \vspace{-0.5em}
    \subfloat[DS2.]{
        \includegraphics[width=0.42\columnwidth]{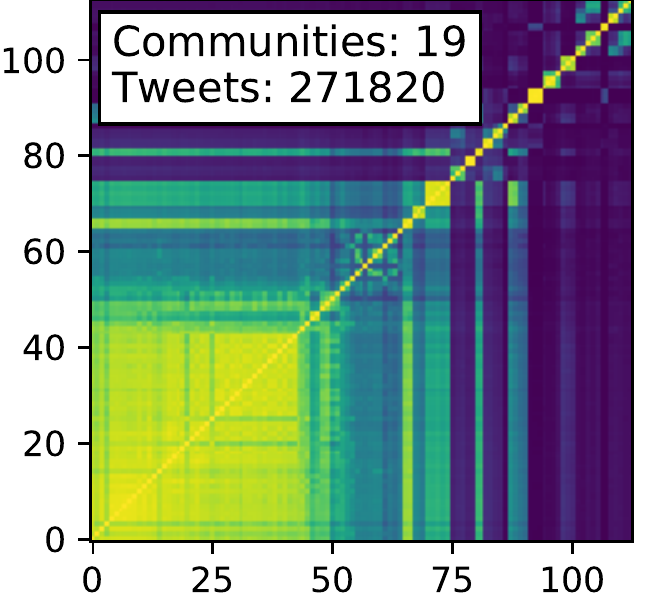}%
        \label{fig:ira_sim_mtx_15m}
    } 
    \subfloat[RANDOM.]{
        \includegraphics[width=0.465\columnwidth]{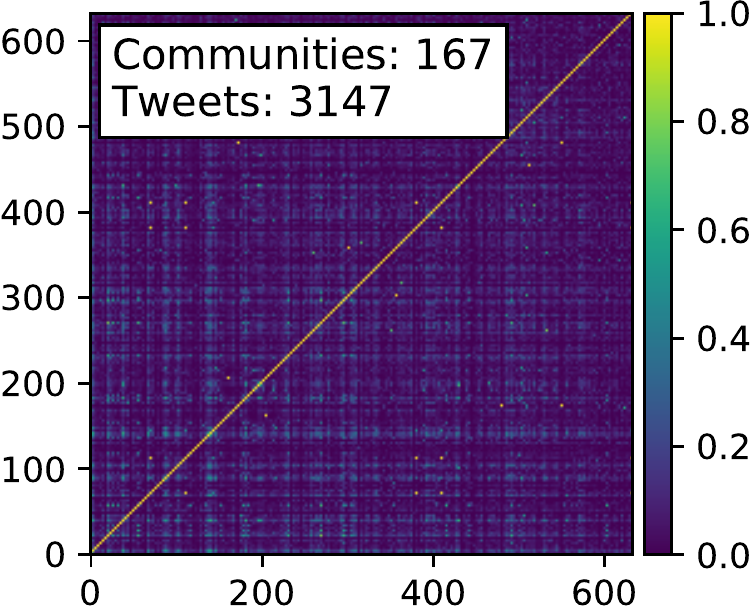}%
        \label{fig:random_sim_mtx_15m}
    }
    \caption{Similarity matrices of content posted by HCC accounts ($\gamma$=15, FSA\_V). Each axis has an entry for each account, grouped by HCC. Each cell represents the similarity between the two corresponding accounts' content, calculated using cosine similarity 
    (yellow = high similarity). Each account's content is represented as a vector of $5$ character n-grams of their combined tweets.}
    \label{fig:hcc_sim_mtxs}
\end{figure}

\textit{Internal consistency.} If HCCs are boosting a message, it is reasonable to assume the content of HCCs members will be more similar
internally that when compared externally, to the content of non-members. 
Treating each HCC member's tweets as a single document, we created a doc-term matrix using $5$ character n-grams for terms, and then compared the members' document vectors using cosine similarity. This approach was chosen for its performance with non-English corpora~\cite{Damashek1995}, and because using individual tweets as documents produced too sparse a matrix. Visualising the similarities between accounts, grouping them by HCC (Figure~\ref{fig:hcc_sim_mtxs}), the HCCs are discernible as being internally similar. 
This method ignores the number of tweets HCCs post, so we can draw no conclusions about connections between HCC size and the internal similarity of their content, though more active HCCs (i.e., with more tweets) are more likely to be similar, through co-occurrence of n-grams.
The RANDOM groupings demonstrated little to no similarity, internal or external, as expected, while the DS2 HCCs demonstrated high internal similarity, as expected of organised accounts over an extended period. 



\begin{figure}[t!]
    \centering
    \subfloat[GT, DS1 and RANDOM.]{
        \includegraphics[width=0.45\columnwidth]{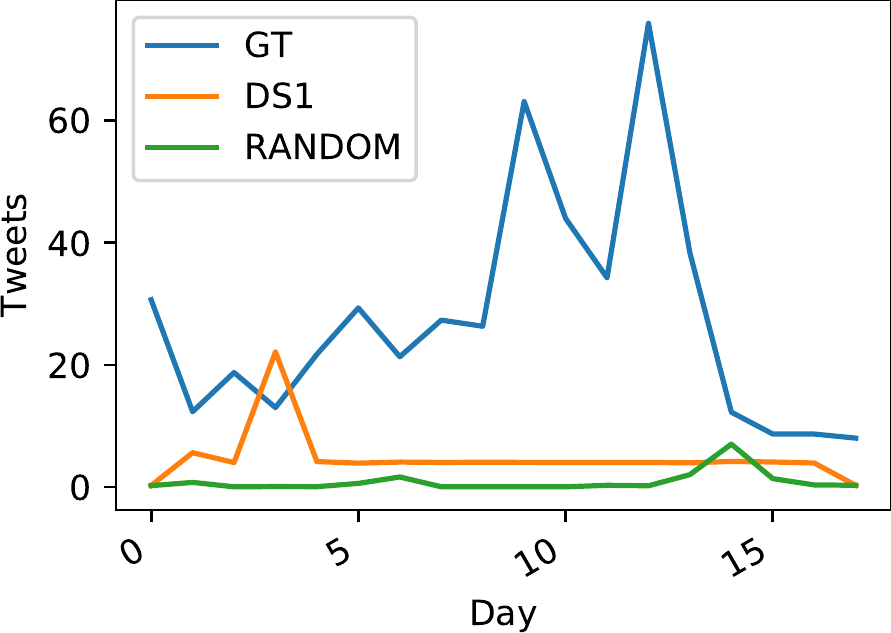}%
        \label{fig:other_hcc_timelines}
    } 
    \subfloat[DS2.]{
        \includegraphics[width=0.47\columnwidth]{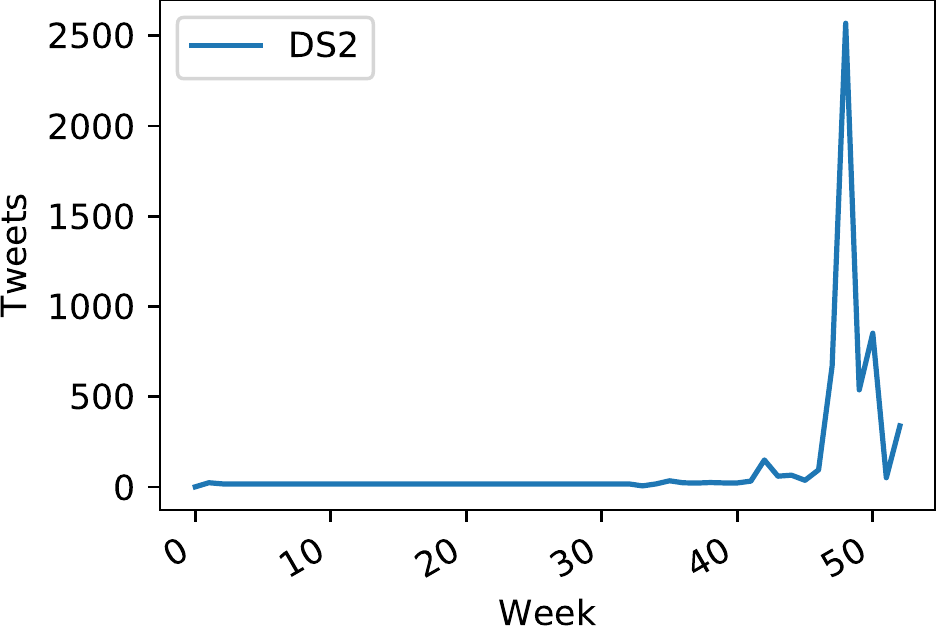}%
        \label{fig:ira_hcc_timeline}
    }
    \caption{Averaged temporal graphs of HCC activities 
    ($\gamma$=$15$, FSA\_V).}
    \label{fig:hcc_timelines}
\end{figure}

\textit{Temporal patterns.} Campaign types exhibit different temporal patterns~\cite{LeeCCS2013campext}. We used the same temporal averaging technique as Lee \emph{et al.}~\cite{LeeCCS2013campext} 
to compare the daily activities of the HCCs found in GT, DS1 and RANDOM 
(Figure~\ref{fig:other_hcc_timelines}) and weekly activities in DS2 (Figure~\ref{fig:ira_hcc_timeline}). The GT accounts were clearly most active at two points prior to the election (around day $15$), during the last leaders' debate and just prior to the mandatory electoral advertising blackout. DS1 and RANDOM HCCs were only consistently active at different times: around the day 3 leaders' debate and on election day, respectively.
Inter-HCC variation may have dragged the mean activity value down, as many small HCCs were inactive each day. Reintroducing FSA's stitching element to FSA\_V may avoid this. 
In DS2, HCC activity increased in the second half of 2016, culminating in a peak around the election, 
inflated by two very active HCCs, both of which used many predominantly benign hashtags over the year. 

\begin{figure}[t!]
    \centering
    \subfloat[GT.]{
        \includegraphics[width=0.455\columnwidth]{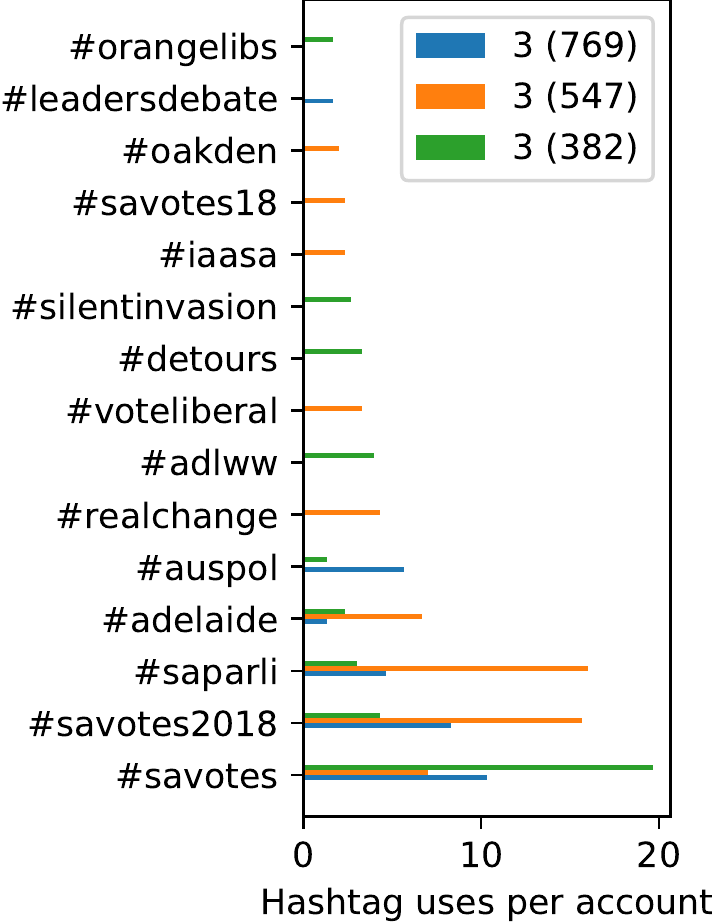}%
        \label{fig:sapol_top_hts_15m}
    } 
    \subfloat[DS1.]{
        \includegraphics[width=0.455\columnwidth]{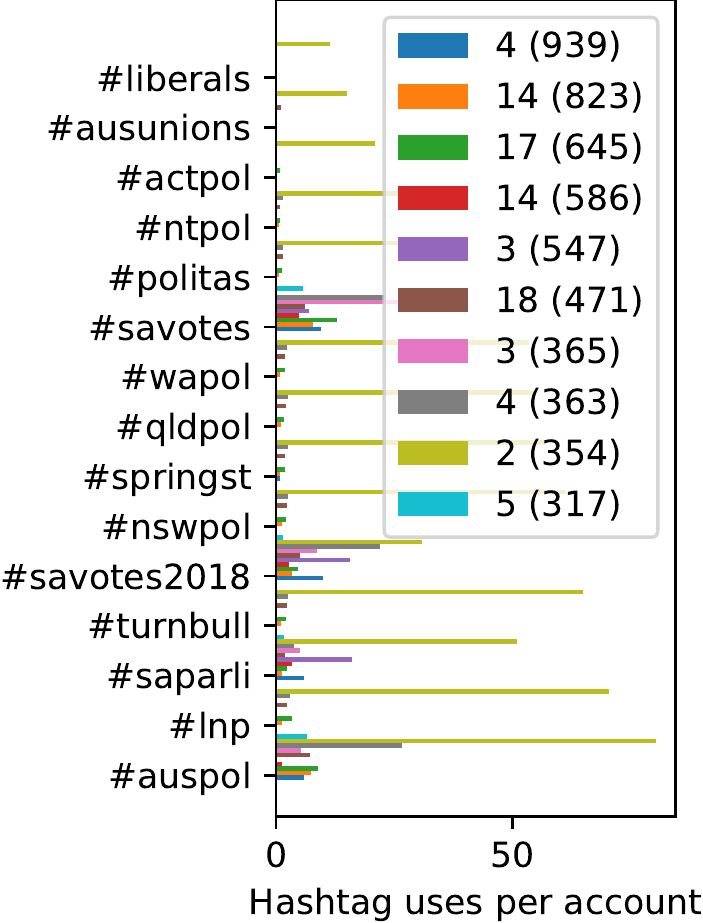}%
        \label{fig:saelec_top_hts_15m}
    } \\ 
    \vspace{-0.5em}
    \subfloat[DS2.]{
        \includegraphics[width=0.775\columnwidth]{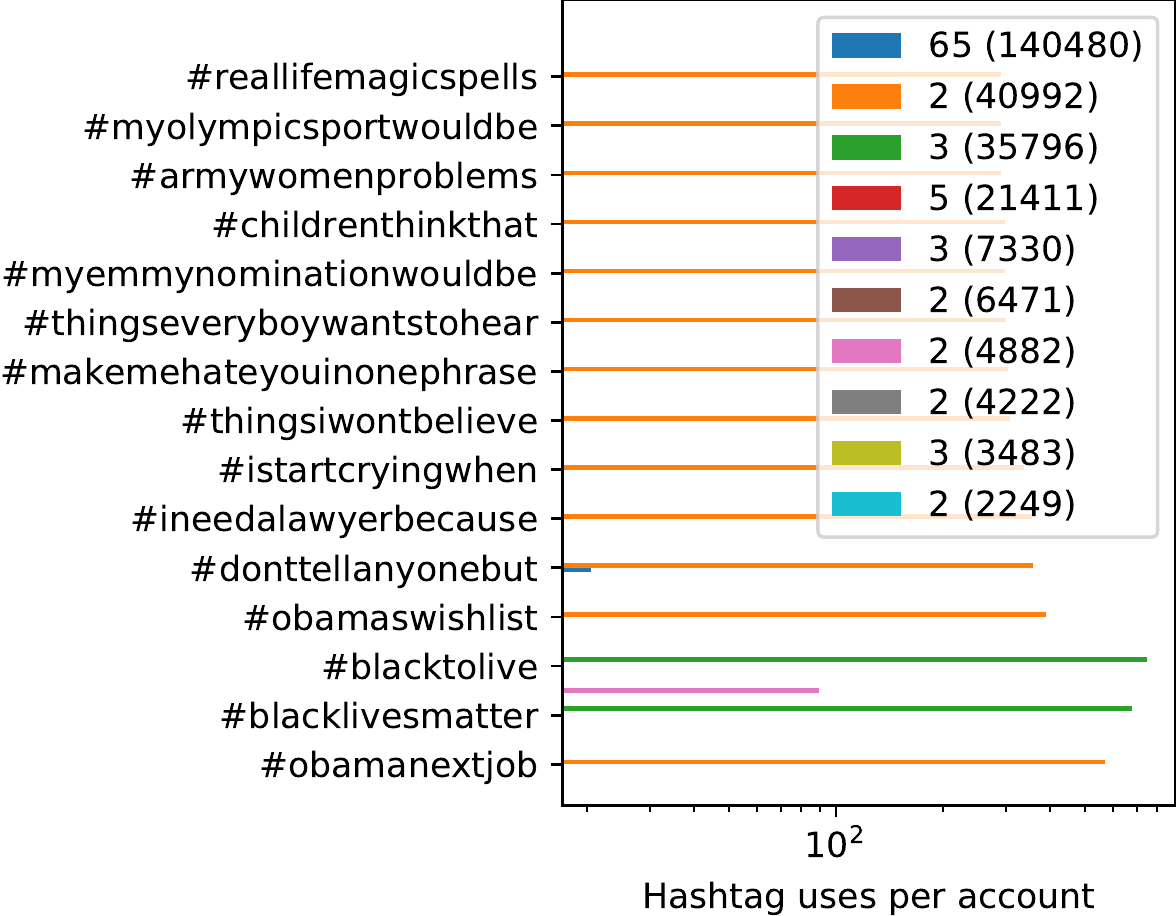}%
        \label{fig:ira_top_hts_15m}
    }
    \caption{Most used hashtags (per account) of the most active HCCs ($\gamma$=15, FSA\_V). The labels indicate member and tweet counts. Not all HCCs used a hashtag often enough to be visible.}
    \label{fig:top_hashtags}
\end{figure}

\textit{Hashtag use.} The most frequent hashtags in the most active HCCs revealed the most in GT. 
It is possible to assign some HCCs to political parties via the partisan hashtags (e.g., {\small \texttt{\#voteliberals}}), although the hashtags of contemporaneous cultural events are also prominent (Figure~\ref{fig:sapol_top_hts_15m}).
DS1 hashtags are all politically relevant, but are dominated by a single small HCC which used many hashtags very often (Figure~\ref{fig:saelec_top_hts_15m}). These accounts clearly attempted to disseminate their tweets through using $1,621$ hashtags in $354$ tweets.
Similarly, DS2 hashtags were dominated by a single HCC (using $41,317$ relatively general hashtags in $40,992$ tweets) and one issue-motivated HCC (Figure~\ref{fig:ira_top_hts_15m}).
Given DS2 covers an entire year, it is unsurprising that the largest HCCs use such a variety of hashtags that their hashtags do not appear on the chart. 

\begin{figure}[t!]
    \centering
    \includegraphics[width=0.45\columnwidth]{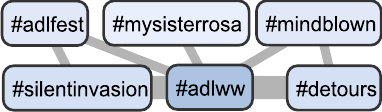}
    \caption{Cluster of hashtags, connected only when they appeared in the same tweet (GT, $\gamma$=15, FSA\_V). Link width = tweet count.}
    \label{fig:sapol_cultural_co-hashtags}
\end{figure}

Analysing co-occurring hashtags 
can help further explore the HCC discussions to determine if HCCs are truly single groups or merged ones. Applied to GT HCC activities (Figure~\ref{fig:sapol_top_hts_15m}), it was possible to delineate subsets of hashtags in use: e.g., one HCC promoted a political narrative in some tweets with {\small \texttt{\#orangelibs}} and discussed cultural events in others with {\small \texttt{\#adlww}} (Figure~\ref{fig:sapol_cultural_co-hashtags}), but was definitely one group.

\textit{Examining the Ground Truth.} 
The importance of having ground truth in context is demonstrated by Keller \emph{et~al.}~\cite{KellerICWSM2017}. 
By analysing the actions of known bad actors in a broad dataset, they could identify not just different subteams within the actors and their strategies, but their effect on the broader discussion. Many datasets comprising only bad actors (e.g., DS2) miss this context.

Considering GT alone, the HCCs identified consist only of members within the same political party, across all values of $\gamma$. 
Some accounts appeared in each window size.  
HCCs of six major parties were identified. 
Examination of these HCCs' content confirmed they were genuine. 
\subsubsection{Focus of connectivity (RQ3)} 

\begin{figure}[t!]
    \centering
    \subfloat[Retweets.]{
        \includegraphics[width=0.45\columnwidth]{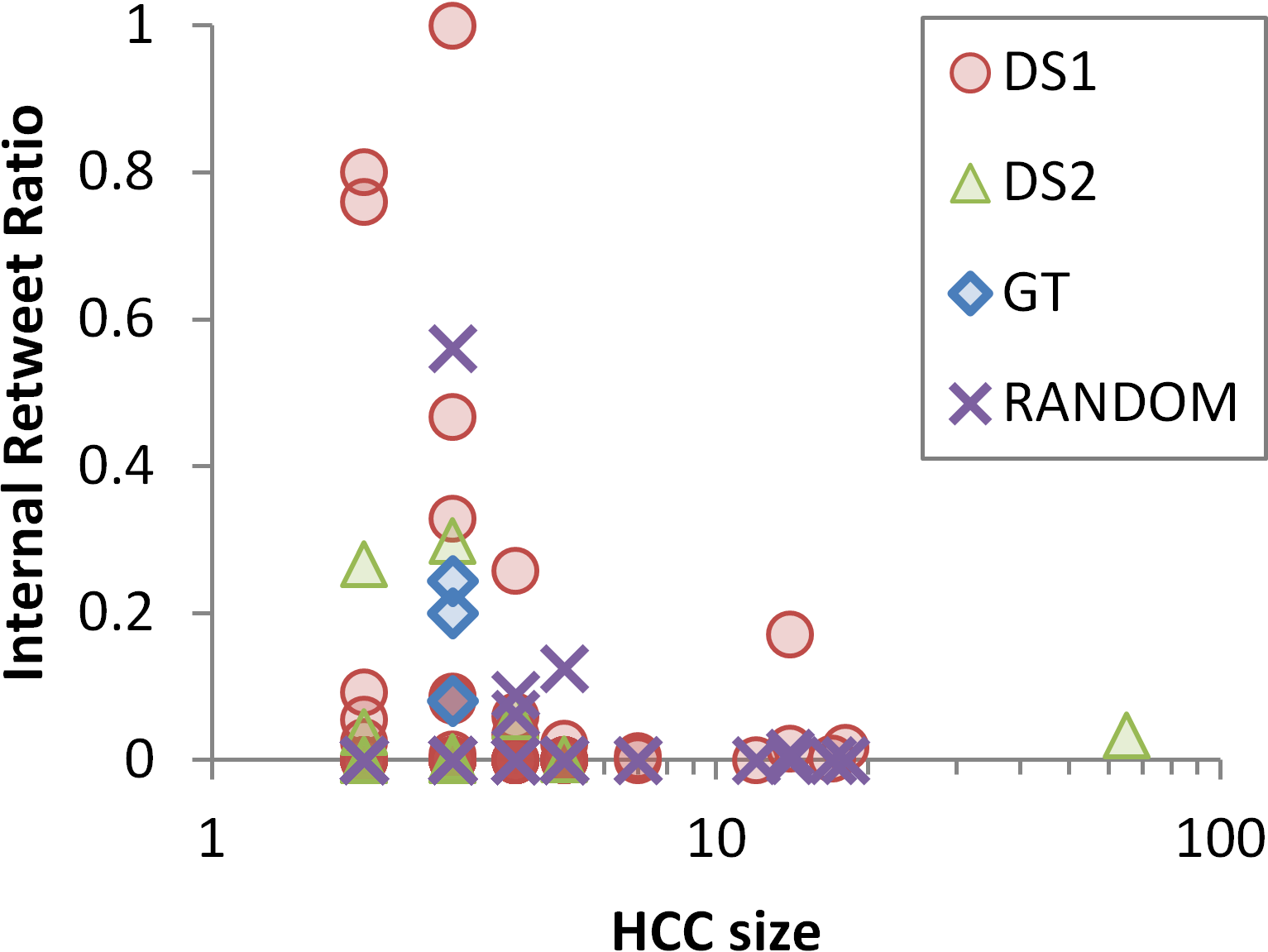}%
        \label{fig:int_rt_ratio_15m_fsa_v}
    } \hfill
    \subfloat[Mentions.]{
        \includegraphics[width=0.45\columnwidth]{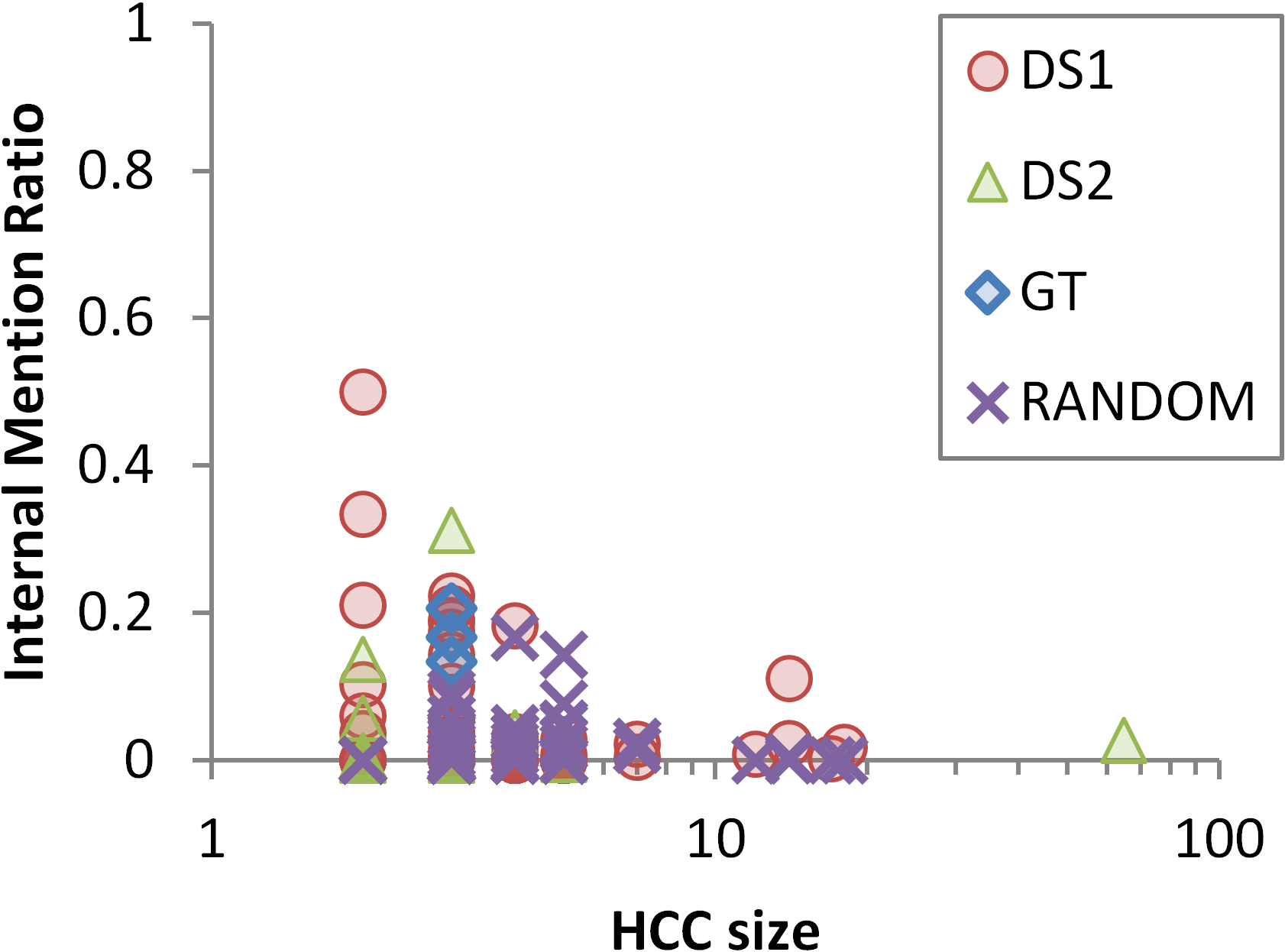}%
        \label{fig:int_m_ratio_15m_fsa_v}
    }
    \caption{The proportions of each HCCs retweets and mentions referring to accounts within the HCC ($\gamma$=15, FSA\_V).}
    \label{fig:hcc_int_ratios}
\end{figure}

Groups that retweet or mention themselves create direct connections; therefore to be covert, it would be sensible to have a low \emph{internal retweet} and \emph{mention ratios} (IRR and IMR, respectively). Figure~\ref{fig:hcc_int_ratios} shows IRR and IMR for the datasets. The larger the HCC size, the greater the likelihood of retweeting or mentioning internally, so it is notable that DS2's largest HCC has IRR and IMR's of around $0$, though even the smaller HCCs have low ratios. Ratios for the smallest HCCs seem largest, possibly due to low numbers of posts, many of which may be retweets or include a mention, inflating the ratios. The hypothesis that political accounts would retweet and mention themselves frequently is not confirmed by these results, possibly because they are retweeting and mentioning official or party accounts outside the HCCs.

\subsubsection{Content variation (RQ4)}

\begin{figure}[t!]
    \centering
    \subfloat[DS1.]{
        \includegraphics[width=0.31\columnwidth]{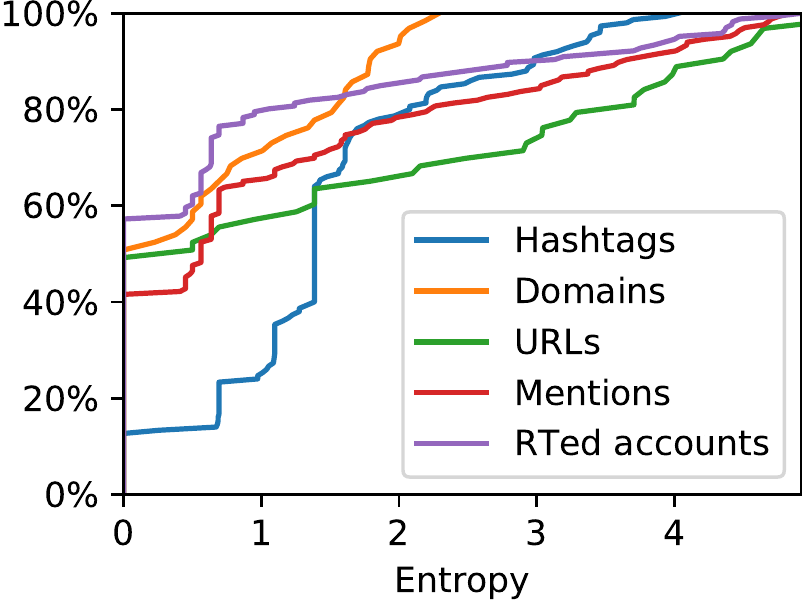}%
        \label{fig:saelec_15m_features_cf}
    } 
    \subfloat[DS2.]{
        \includegraphics[width=0.31\columnwidth]{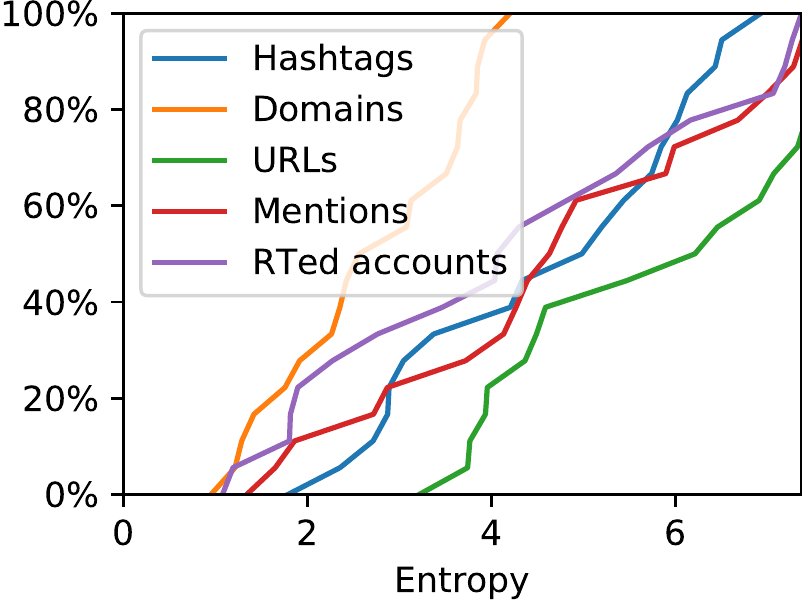}%
        \label{fig:ira_15m_features_cf}
    } 
    \subfloat[RANDOM.]{
        \includegraphics[width=0.31\columnwidth]{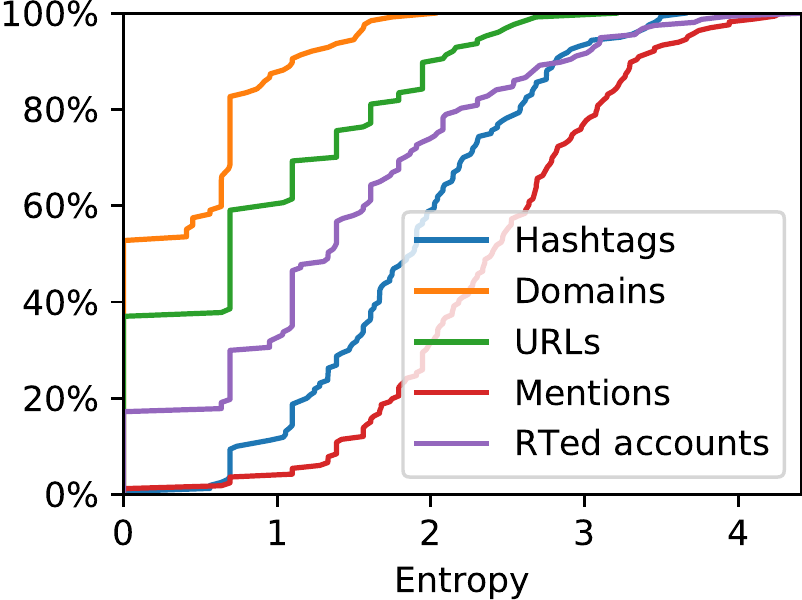}%
        \label{fig:random_15m_features_cf}
    }
    \caption{Cumulative frequency of HCCs' entropy scores for five tweet features, comparing DS1 and DS2 with RANDOM ($\gamma$=$15$, FSA\_V). Feature variation increases moving right on the x axis.}
    \label{fig:features_cfs}
\end{figure}

Highly coordinated reposting involved reusing the same content frequently, resulting in low feature variation (e.g., hashtags, URLs, mentioned accounts), which can be measured as entropy~\cite{CaoCLGC2015urlsh}. A frequency distribution of each HCC's use of each feature type was used to calculate each entropy score. Low feature variation corresponds to low entropy values. As per~\cite{CaoCLGC2015urlsh}, we compared the entropy of features used by DS1 and DS2 HCCs to RANDOM ones (Figure~\ref{fig:features_cfs}). Entries for HCCs which did not use a particular feature are omitted, as their scores would inflate the number of groups with $0$ entropy. 
Many of DS1's small HCCs used only one of a particular feature, resulting in an entropy score of $0$ (Figure~\ref{fig:saelec_15m_features_cf}). In contrast, DS2's fewer HCCs have higher entropy values (Figure~\ref{fig:ira_15m_features_cf}), likely because they were active for longer (over $365$, not $18$, days) and had more opportunity to use more feature values. 
The majority of HCCs used few hashtags and URL domains, which is expected as the dominating domain is \emph{twitter.com}, embedded in all retweets. Compared to the RANDOM HCCs (Figure~\ref{fig:random_15m_features_cf}), DS1 HCCs had lower variation in all features, while the longer activity period of DS2 resulted in distinctly different entropy distributions.

\subsubsection{Multiple criteria: \emph{Bully}ing}

Some strategies can involve a combination of actions. Behaviours that contribute to \emph{Bully}ing by dogpiling, for example, include joining conversations started by the target's posts and mentioning the target repeatedly, within a confined timeframe. As DS1 included all replied to tweets, we investigated it inferring links via co-mentions and co-convs with a window size of $10$ minutes, and FSA\_V with $\theta$=$0.001$, having maximised the ratio of MEW to HCC size. Of $142$ HCCs discovered, the largest had five accounts and most only had two. Only $32$ had more than ten inferred connections, but five have more than $1,000$. These heavily connected accounts, after deep analysis, were simply very active Twitter users who engaged others in conversation via mentions, which outweighed the more strict co-conv criterion of participants \emph{reply}ing into the same conversation reply tree. 

\begin{figure}[t!]
    \centering
    \includegraphics[width=0.99\columnwidth]{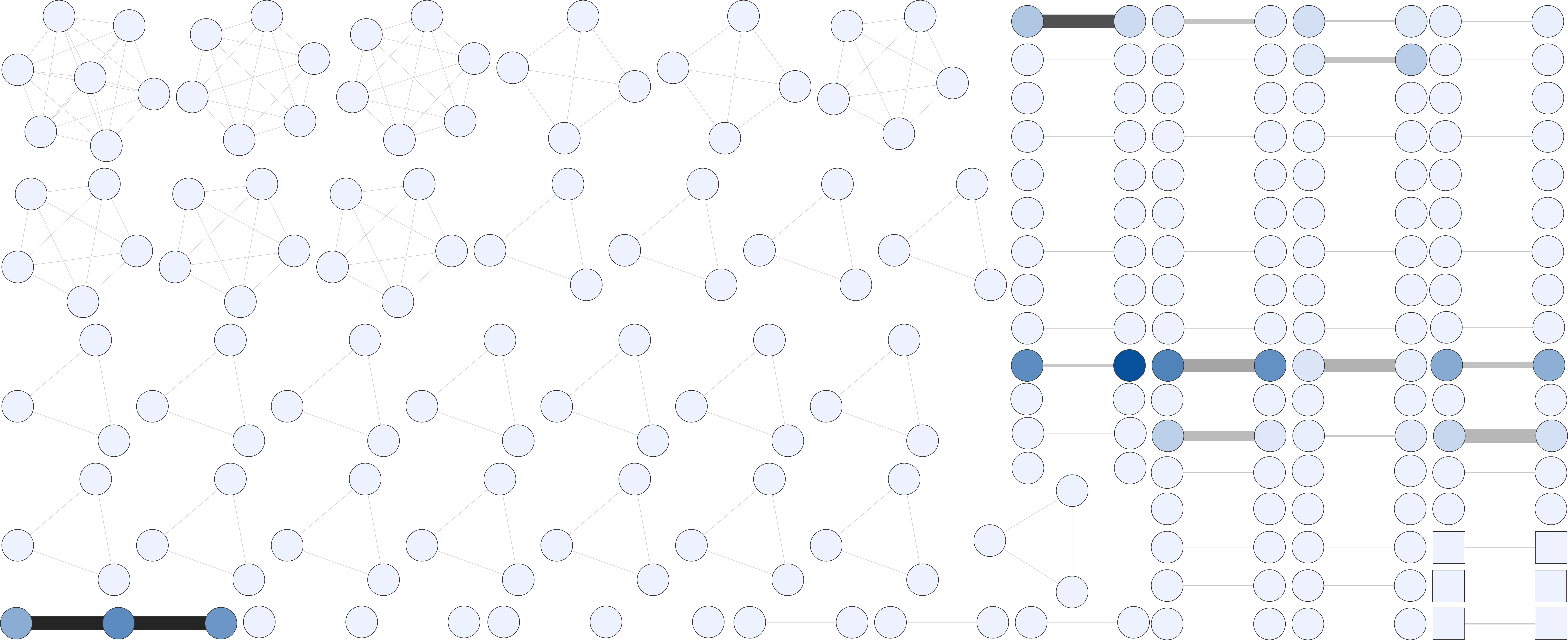}%
    \caption{While searching for \emph{Bully}ing behaviour in DS1, these are HCCs of accounts found engaging in co-mentions (circles) and co-mentions plus co-convs, i.e., engaged in both (square vertices in bottom right) ($\gamma$=360, FSA\_V, $\theta$=$0.01$). Edge thickness and darkness = inferred connections (darker = more). Vertex colour = tweets posted by that account (darker = more). Created with \emph{visone} (https://visone.info).}
    \label{fig:ds1_bully_360m}
\end{figure}

A larger window size was considered ($\gamma$=$360$) in case co-conv interactions were more prevalent. FSA\_V ($\theta$=$0.01$) exposed little further evidence of co-conv (Figure~\ref{fig:ds1_bully_360m}), finding $98$ small HCCs again dominated by co-mentions, not many of which had more than one inferred connection, implying most links were incidental; FSA\_V did not filter these out.

This provides an argument for a more sophisticated approach to combining LCN edge weights for analysis, instead of Equation (\ref{eq:lcn_uv_combined_weight}), and that FSA\_V could be modified to better balance HCC size and edge weight. Furthermore, it is likely that bullying accounts will not just co-mention accounts frequently, but have low diversity in the accounts they co-mention, i.e., they repeatedly co-mention a small number of accounts. That nuance is not explored here.

\subsubsection{HCC inter-relationships}

\begin{figure}[t!]
    \centering
    \includegraphics[width=0.9\columnwidth]{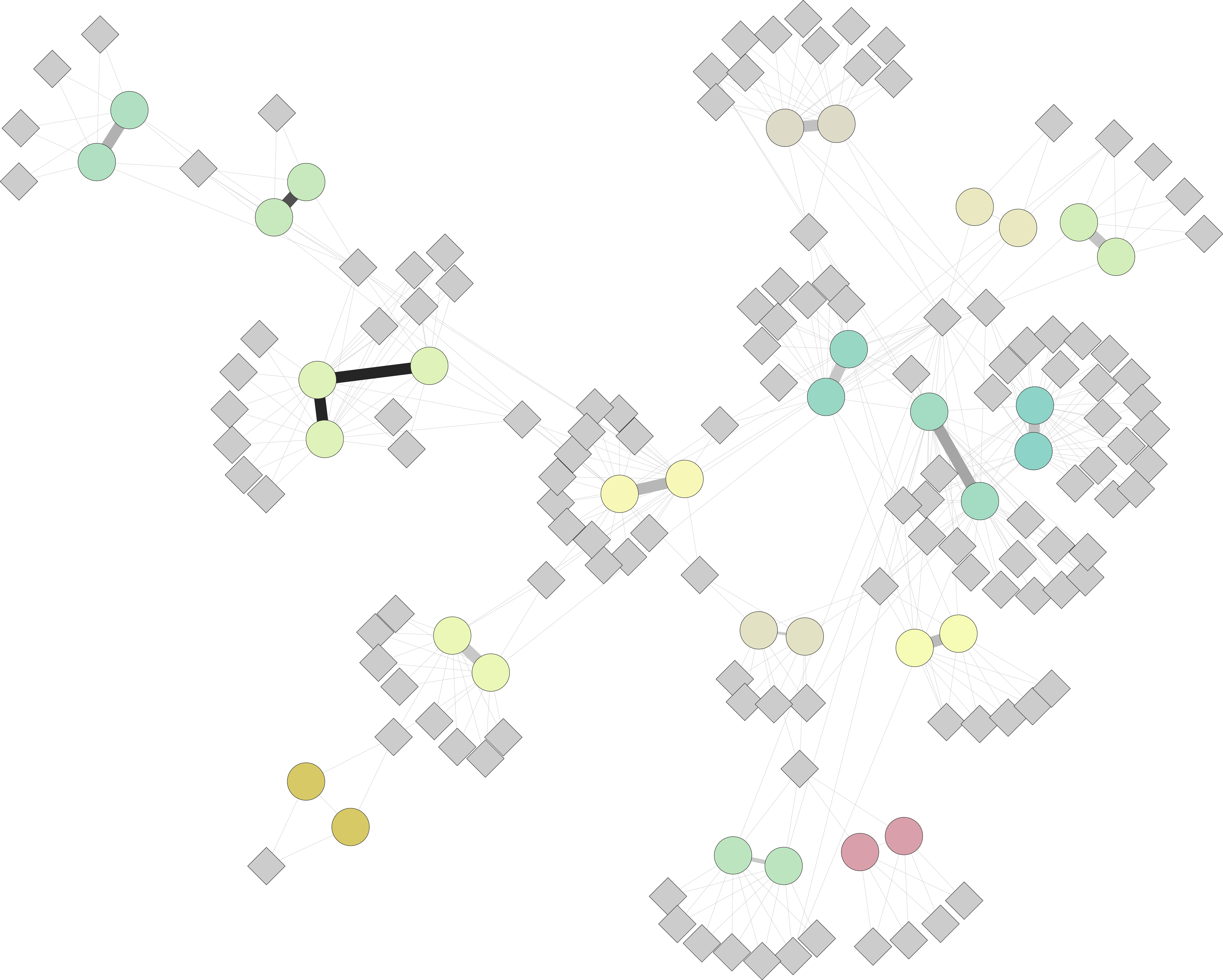}%
    \caption{A graph of DS1 HCC accounts (circle vertices) connected to the accounts they mention or conversations they join (diamonds). Accounts in the same HCC share colours. Clear communities surrounding HCCs indicate who they converse with, and which conversants are co-mentioned by mulitple HCC accounts. Created with \emph{visone} (https://visone.info).}
    \label{fig:ds1_bully_360m_exp}
\end{figure}

Introducing vertices to represent the \emph{reasons} HCC accounts are connected (e.g., who they co-mention, or conversations they join) shows how the HCCs inter-relate. Figure~\ref{fig:ds1_bully_360m_exp} shows the largest component after such expansion was conducted on the HCCs in Figure~\ref{fig:ds1_bully_360m}. HCC accounts (circles) share colours and the distribution of the reasons for their connection (diamonds) show which are unique to HCCs and which are shared. Heavy links between HCC accounts with few adjacent reason vertices imply these are accounts mentioned many times.

\subsubsection{\emph{Boost}ing accounts, not tweets}

\begin{figure}[t!]
    \centering
    \includegraphics[width=0.99\columnwidth]{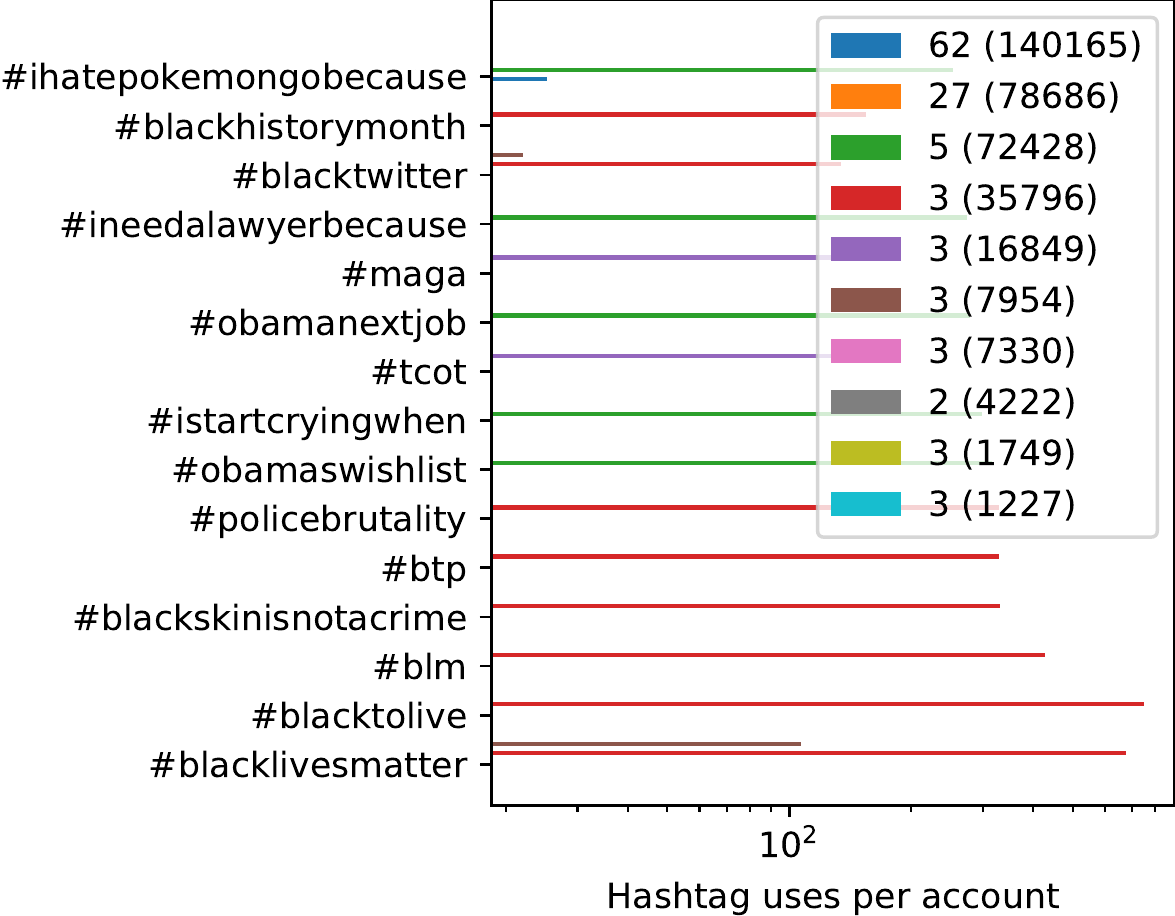}%
    \caption{Most used hashtags (per account) of the most active HCCs boosting accounts (FSA\_V, $\gamma$=$15$). The labels indicate member and tweet counts. Not all HCCs used a hashtag often enough to be visible.}
    \label{fig:top_hts_boost_users}
\end{figure}

It is possible to \emph{Boost} an account rather than just a post. Returning to DS2, we sought HCCs from accounts retweeting the same account (FSA\_V, $\gamma$=$15$), and found that the hashtag use revealed further insights (Figure~\ref{fig:top_hts_boost_users}). No longer does one HCC dominate the hashtags. Instead clear themes are exhibited by different HCCs, but again, they are not the largest HCCs. The red HCC uses {\small \texttt{\#blacklivesmatter}}-related hashtags, while the purple HCC uses pro-Republican ones ({\small \texttt{\#maga}} and {\small \texttt{\#tcot}}), and the green HCC is more general. Given the number of tweets these HCCs posted over 2016 (at least $16,849$), it is clear they concentrated their messaging on particular topics, some politically charged.


\section{Conclusion}

As online influence operations grow in sophistication, our automation and campaign detection methods must also expose the accounts covertly engaging in ``orchestrated activities''~\cite{GrimmeAA2018perspectives}. We have described several coordination strategies, their purpose and execution methods, and demonstrated a novel pipeline-based approach to finding sets of accounts engaging in such behaviours in two relevant datasets. Using discrete time windows, we temporally constrain potentially coordinated activities, successfully identifying groups operating over various timeframes.
The analysis of HCC evolution, improvement of HCC extraction techniques and investigation of near real time processing provide opportunities for future research in this increasingly important field.





\bibliographystyle{IEEEtran}
\bibliography{IEEEabrv,refs}

\end{document}